\documentclass[reprint,prb,aps,epsfig,longbibliography,superscriptaddress,twocolumn]{revtex4-2}
\usepackage{graphicx}
\usepackage[unicode=true,colorlinks=true]{hyperref}
\hypersetup{linkcolor=blue,citecolor=blue,urlcolor=blue}
\usepackage{dcolumn}
\usepackage{bm}
\makeatletter
\usepackage{color, colortbl}
\definecolor{Gray}{gray}{0.9}

\newcommand{\Rmnum}[1]{\expandafter\@slowromancap\romannumeral #1@}

\begin{document}

\title{Temperature dependent equilibration of spin orthogonal quantum Hall edge modes}

\begin{abstract}
Conductance of the edge modes as well as conductance across the co-propagating edge modes around the $\nu$ = 4/3, 5/3 and 2 quantum Hall states are measured by individually exciting the modes. Temperature dependent equilibration rates of the outer unity conductance edge mode are presented for different filling fractions. We find that the equilibration rate of the outer unity conductance mode at $\nu$ = 2 is higher and more temperature sensitive compared to the mode at fractional filling 5/3 and 4/3. At lowest temperature, equilibration length of the outer unity conductance mode tends to saturate with lowering filling fraction $\nu$ by increasing magnetic field $B$. We speculate this saturating nature of equilibration length is arising from an interplay of Coulomb correlation and spin orthogonality.
\end{abstract}

\author{Tanmay Maiti}
\email{tanmay.maiti@saha.ac.in}
\affiliation{Saha Institute of Nuclear Physics, HBNI, 1/AF Bidhannagar, Kolkata 700064, India}
\author{Pooja Agarwal}
\affiliation{Saha Institute of Nuclear Physics, HBNI, 1/AF Bidhannagar, Kolkata 700064, India}
\author{Suvankar Purkait}
\affiliation{Saha Institute of Nuclear Physics, HBNI, 1/AF Bidhannagar, Kolkata 700064, India}
\author{G J Sreejith}
\affiliation{Indian Institute of Science Education and Research, Pune 411008, India}
\author{Sourin Das}
\affiliation{Department of Physical Sciences, IISER Kolkata, Mohanpur, West Bengal 741246, India}
\author{Giorgio Biasiol}
\affiliation{Istituto Officina dei Materiali CNR, Laboratorio TASC, 34149 Trieste, Italy}
\author{Lucia Sorba}
\affiliation{NEST, Istituto Nanoscienze-CNR and Scuola Normale Superiore, Piazza San Silvestro 12, I-56127 Pisa, Italy}
\author{Biswajit Karmakar}
\affiliation{Saha Institute of Nuclear Physics, HBNI, 1/AF Bidhannagar, Kolkata 700064, India}

\maketitle
\maketitle
\section{INTRODUCTION}
Quantum Hall (QH) systems formed the first examples of topological insulators, where a set of gapless, topologically protected edge modes carry the current around a bulk region that is gapped due to an interplay of the applied magnetic field and interaction. Robustness of the QH systems has allowed extensive theoretical and experimental investigations of the detailed chiral edge transport revealing rich physics arising along the one dimensional boundary \cite{RevModPhys.75.1449,MacDonald1990,Beenakker1990,Meir1994,Sabo2017,Lin2019}. Though the bulk QH state fixes the total charge conductance along the boundary, the confinement potentials and electronic interactions can reconstruct the edge modes affecting the details of the current distribution\cite{Beenakker1990,Wen1994,Johnson1995,Wan2002,Wang2013,Wan2003,Joglekar2003,Yang2003}. Robustness and coherence of reconstructed edge modes \cite{Ronen2018,Nakamura2019} can have implications in investigation of quantum interferrometry \cite{McClure2012,Ofek2010,Park2015}, braiding statistics \cite{Nakamura2020,Halperin1984,Kim2006,Arovas1984}, and in QH circuit designs for quantum electronics applications \cite{Ji2003,Neder2007}.  Weakly equilibrating fractional conductance modes have been realized around $\nu=1$ \cite{Kouwenhoven1990,Bhattacharyya2019} as well as around $\nu=2/3$ \cite{Sabo2017,Nosiglia2018,Wang2013,Gross2012} states, where larger equlibration lengths are achieved in the high magnetic field limit\cite{Maiti2020}. Characterization of the equilibration processes is thus a question of active interest.

In this work, we focus on equilibration of the spin orthogonal edge modes occuring around states at filling fractions higher than 1 namely at $\nu=4/3,5/3$ and $2$. The $\nu = 4/3$ state is fully spin polarized in Si/SiGe heterostructures \cite{Lai2004}. A spin polarization transition can occur in 2D hole gas\cite{Davies1991,Rodgers1993,Muraki1998} as well as in 2D electron gas \cite{Clark1989} embedded in GaAs/AlGaAs heterostructures under tilted magnetic field induced excess Zeeman splitting. The edge of the spin unpolarized $\nu = 4/3$ QH state in GaAs/AlGaAs heterostructure has two co-propagating spin orthogonal charge modes with conductance $1$ and $1/3$, \cite{Altimiras2012,Lin2019} with equilibration lengths measured upto of a few hundred micrometers. Edge structure of integer QH state at $\nu=2$ is well understood,\cite{Halperin1982,Muller1992,Kang2000,Karmakar2011} where two unity conductance spin orthogonal co-propagating edge modes carry the current. Scattering between the spin orthogonal edge modes can occur through spin flip processes assisted by the dynamics of nuclear spins \cite{Machinda2002,Yang2017,Kane1992,Wald1994,Dixon1997,Deviatov2004}. 
I-V spectroscopy of the QH states at filling fraction $\nu$ = 4/3,5/3, 2 and 3 shows that conductance across the copropogating edge modes is enhanced above a inter-edge mode threshold bias $V_{th}$ and the threshold voltage $V_{th}$ increases with decreasing filling fraction $\nu$ \cite{Deviatov2008}. Above the threshold voltage, conductance across the edge modes reaches the equilibration value for integer filling fraction $\nu = 2$ and 3. In contrast, for fractional filling $\nu = 5/3$ and 4/3 the conductance across the modes is below the corresponding equilibrium value above the threshold voltage. Above the threshold voltage $V_{th}$ inter-edge equilibration become faster because of the flat band scenario \cite{Deviatov2008}. As a consequence, it is difficult to estimate the equilibration length/rate of the edge modes at high imbalance. In this paper, we intend to study the equilibration between the spin orthogonal edge modes in the linear transport regime at filling fraction $\nu$ = 4/3, 5/3 and 2.

In this article, we study conductance of the edge modes and conductance across the co-propagating edge modes around the $\nu=4/3,\,5/3$ and $2$ QH states by individually exciting the modes. We measure the length scales over which the outer unity conducatnce mode (present in all the QH states under study) equilibrates with the inner modes of conductance 1, 2/3 and 1/3 for the bulk filling $\nu = 2, 5/3$ and 4/3, respectively. We find that the equilibration rate of the outer unity conductance mode at $\nu$ = 2 is higher and more temperature sensitive than that of fractional filling 5/3 and 4/3. At lowest temperature the equilibration length of the outer unity conductance mode shows saturating behaviour with increasing magnetic field $B$ i.e. decreasing bulk filling fraction $\nu$. The observation of significantly larger equilibration length in the case of filling 4/3 and 5/3 relative to the $\nu$ = 2 case potentially indicating a slower equilibration between an integer and fractional mode.

\section{DEVICE DESCRIPTION AND MEASUREMENT PROCEDURE}
\begin{figure}[h!]
\includegraphics[width= 7.5 cm]{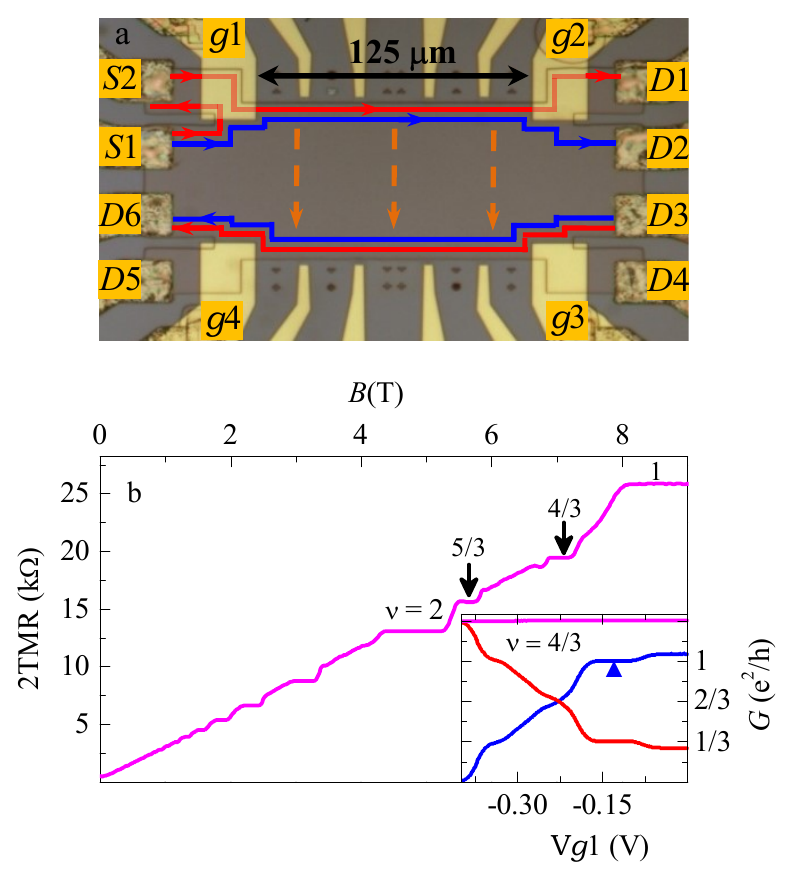}
\centering \caption[ ] 
{\label{fig:device diagram} (a) Optical image of the device used for transport measurement where relevant edge mods for bulk filling fractions $\nu=4/3, 5/3, 2$ are shown. Outer red channel is unity conductance mode and blue channel represents inner mode. Edge modes are separately contacted by setting the filling fraction beneath the two top gates at unity $\nu_1=\nu_2=1$. (b) Two terminal magneto-resistance trace taken at base temperature. Inset shows $g$1 gate characteristics at filling fraction $\nu=4/3$ keeping $g2$ at pinch off condition, where blue line represents transmittance conductance ($S1 \rightarrow S2$), red line represents reflected conductance ($S1 \rightarrow D2$) and magenta line represents total conductance.}
\end{figure}

Experiments are carried out on a modulation doped GaAs/AlGaAs heterostructure, in which the two-dimensional electron gas (2DEG) resides at the GaAs/AlGaAs heterointerface located $100 \ {\rm nm}$ below the top surface. Figure \ref{fig:device diagram}(a) shows the device structure, where eight Ohmic contacts $S1,\,S2,\,D1,\,D2,\,D3,\,D4,\,D5,\,D6$ are defined for current injection and detection and four top gates $g1,\,g2,\,g3,\,g4$ are used to tune the filling fraction in the mesa underneath the gates. A customized pre-amplifier SR555 \cite{Maiti2020} is deployed at $S2$ contact to measure output current and for application of ac voltage excitation simultaneously. The device is mounted in a dilution refrigerator equipped with a 14~T superconducting magnet at a base temperature 7~mK, where electron temperature achieved is about 30~mK. Carriers are injected by illumination with a GaAs LED in the sample at $3$ K and these carriers are persistent at low temperatures \cite{nathan1986persistent}. The carrier density and mobility of the sample become $n \sim 2.27\times10^{11} \ {\rm cm^{-2}}$ and $\mu \sim 4\times10^6 \ {\rm cm^2/Vs}$, respectively after light illumination. Two terminal magnetoresistance (2TMR)  [Fig. \ref{fig:device diagram}(b)] is measured to find the location of the QH states namely $\nu=4/3,\,5/3,\,2$ along the magnetic field axis at the base temperature. For our transport experiments at filling $\nu = 4/3,\,5/3,\,2$ the magnetic fields are set at $7.1,\,5.69,\,4.9$ T (as indicated in Fig. \ref{fig:device diagram}(b)),  respectively. The top gates $g3$ and $g4$ are kept at pinch-off condition by applying a negative voltage bias of $Vg3 = Vg4 = -0.450 $ V through out the experiment.

We set the magnetic field at $7.1$ T to carry out top gate $g1$ characteristics at a bulk filling fraction $\nu = 4/3$ keeping the top gate $g2$ in pinch-off condition. Transmitted and reflected conductance are measured between $S1 \rightarrow S2$ and $S1 \rightarrow D2$, respectively by varying the top gate voltage $Vg1$ as shown in inset of Fig. \ref{fig:device diagram}(b).The transmitted conductance (blue curve) shows a plateau at unit conductance, confirming the formation of an integer QH state of filling $\nu_1 = 1$ beneath the top gate $g1$ within a gate voltage range of $-0.168$ to $-0.099$ V. Red curve represents reflected conductance and magenta curve shows total conductance. The total conductance remains constant at $4/3$ throughout the gate voltage $Vg1$ scan confirming conservation of the current. A similar gate characteristic is also observed for the top gate $g2$ at a bulk filling $\nu = 4/3$. From the gate characteristics, we can determine the gate voltage range in which an integer QH state of filling unity is formed beneath the gates $g1$ and $g2$. Similarly, gate voltage ranges of top gate $g1$ and $g2$ are found to set the filling fraction unity $\nu_1=\nu_2=1$ beneath the gates at bulk filling fraction $\nu=5/3$ and $2$.

\section{EXPERIMENTAL RESULTS AND ANALYSIS}
Upon setting $\nu_1=\nu_2=1$ beneath the gates, we separately contact the edge modes for selective current injection and detection \cite{Karmakar2011,Wurtz2002} in the $\nu = 4/3,\,5/3,\,2$ QH states. In this device, outer unity conductance mode (red line) connects $S2$ to $D1$ and $S1$ is connected to $D2$ by the inner edge mode (blue line) with conductance $1/3, 2/3, 1$ for filling fraction $\nu=4/3,\,5/3,\,2$, respectively \cite{Beenakker1990,Lin2019} as shown in Fig. \ref{fig:device diagram}(a). To understand the equilibration between these edge modes, we perform the temperature dependent transport measurements from $30$ mK to only $500$ mK preventing thermal degradation of the sample - loss of carriers and reduction of mobility. During temperature dependence experiment, $S1$ and $S2$ are excited with $V_{S1}=25.8 \mu V$ ($17$ Hz) and $V_{S2}=25.8 \mu$V ($26$ Hz) and current at $D1$ and $D2$ are measured with the two frequency windows and back-scattered current ($17$ Hz) is measured at $D6$. With this excitation voltage, we work in the linear transport regime. We define the measured quantities  $^{\nu }G_{S \rightarrow D}$ which denote two terminal conductance (TTC) between a source $S$ and a  drain $D$ at bulk filling fraction $\nu$, while filling fractions $\nu_1 = \nu_2 = 1$ is maintained under the gates. The TTC  $^{\nu}G_{S \rightarrow D}$ can also be expressed in terms of transmission probabilities of the co-propagating edge modes (see Appendix A).

\begin{figure}[h!]
\includegraphics[width=7.5 cm]{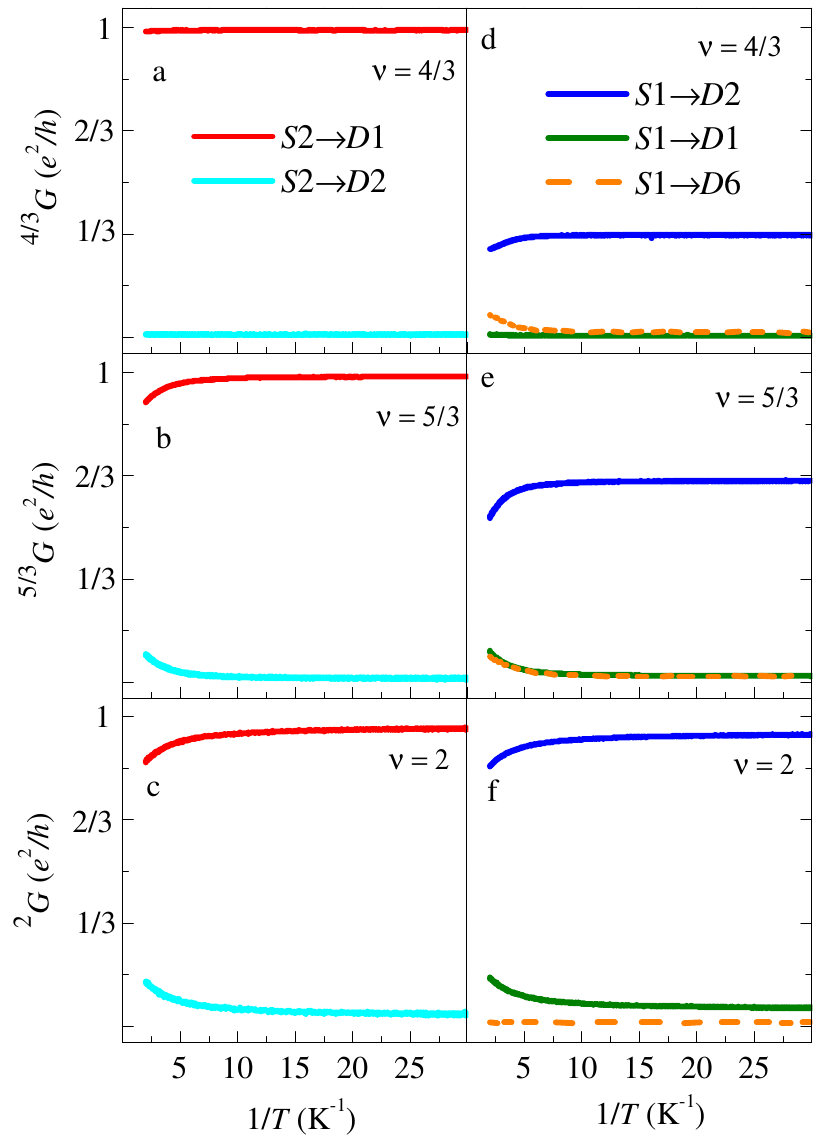}
\centering \caption[ ] 
{\label{fig:temp} Two terminal conductances as a function of $1/T$, when S2 is excited with $25.8~\mu$V at $26$ Hz for bulk filling fractions (a) $\nu=4/3$, (b) $\nu=5/3$ and (c) $\nu=2$. Two terminal conductances as a function of $1/T$  when $S1$ is excited with $25.8~\mu$V at $17$ Hz for (d) $\nu=4/3$, (e) $\nu=5/3$ and (f) $\nu=2$.}
\end{figure}

Temperature dependent TTCs are presented in Fig. \ref{fig:temp}. At filling fraction $\nu=4/3$, TTCs between contacts $S2 \rightarrow D1$, and $S2 \rightarrow D2$ are shown in Fig. \ref{fig:temp}(a) by the red and cyan curves, respectively. With increasing temperature upto $500$ mK, we see that TTC $^{4/3}G_{S2 \rightarrow D1}$ stays fixed to $1$ and zero conductance is measured for TTC $^{4/3}G_{S2 \rightarrow D2}$, indicating no measurable equilibration between the outer unity conductance mode and inner $1/3$ conductance mode upto $500$ mK over a propagation length of $l= 125 \,\mu$m. Similar temperature dependent experiments for filling fraction $\nu=5/3$ and $2$ shows that with increasing temperature, $^{5/3}G_{S2 \rightarrow D1}$ and  $^{2}G_{S2 \rightarrow D1}$ decrease [red curve of Fig. \ref{fig:temp}(b) and \ref{fig:temp}(c)] while $^{5/3}G_{S2 \rightarrow D2}$ and  $^{2}G_{S2 \rightarrow D2}$ increase [cyan curve of Fig. \ref{fig:temp}(b) and \ref{fig:temp}(c)]. Compensating nature of the conductances confirm the conservation of current. At higher temperature, the outer unity conductance mode equilibrates with inner 2/3 and 1 conductance mode for bulk filling fraction $\nu=5/3$ and $2$, respectively. Suppression of inter-mode scattering over propagation length of $\l = 125 \,\mu$m at lowest temperature is evident in Figs. \ref{fig:temp}(a-c).

The TTC $^{4/3}G_{S1 \rightarrow D2}$ decreases with increasing temperature as shown in Fig. \ref{fig:temp}(d)  (blue curve) while $^{4/3}G_{S1 \rightarrow D1}$ (olive curve) remains fixed to zero due to absence of equilibration upto $500$ mK as also seen for $^{4/3}G_{S2 \rightarrow D2}$ in Fig. \ref{fig:temp}(a). The TTC for the current reaching at $D6$ from $S1$ increases with increasing temperature [dashed orange curve of Fig. \ref{fig:temp}(d)], indicating back-scattering of the inner $1/3$ edge mode into the oppositely moving edge channel across the bulk [Fig. \ref{fig:device diagram}(a)]. For filling fraction $\nu=5/3$ in the bulk, TTC $^{5/3}G_{S1 \rightarrow D2}$ ($^{5/3}G_{S1 \rightarrow D1}$)  decreases (increases) with increasing temperature as shown in Fig. \ref{fig:temp}(e) by blue (olive) curve and  the back-scattered current at $D6$ also increases [dashed orange curve of Fig. \ref{fig:temp}(e)] with increasing temperature. The observation indicates simultaneous equilibration of co-propagating modes and backscattering of the 2/3 mode with increasing temperature. At a bulk filling fraction $\nu=2$, decrease of $^{2}G_{S1 \rightarrow D2}$ is fully compensated  by increase of  $^{2}G_{S1 \rightarrow D1}$, and no current reaches at $D6$ [Fig. \ref{fig:temp}(f)], which confirms incompressibility of QH state at $\nu=2$ within the range of temperature variations. At filling fraction $\nu=4/3$ and $5/3$, the sub-Kelvin bulk gaps (see Appendix B) originate from the coulomb interaction, while the bulk gap at filling fraction $\nu=2$ is the Landau gap $\hbar\omega_c$ ($\omega_c$ - cyclotron frequency). Hence, break-down of the QH state at $\nu = 2$ is not observed, as the cyclotron gap $\hbar\omega_c$ is much larger than the maximum applied thermal excitation $kT$ ($T = 500$ mK).

To quantify the temperature dependence of equilibration process between the edge modes, we define equilibration length $l_r$ of the outer integer edge mode, where $1/l_r$ is the rate of charge transfer from outer to inner mode.  Corresponding TTC of the outer mode connecting $S2$ to $D1$ can be written as \cite{Muller1992,Wurtz2002,Maiti2020,Takagaki1994,Paradiso2011}  
\begin{equation}
^{2}G_{S2 \rightarrow D1} = \frac{1}{2}[1+ e^{-2l/l_r}],
\label{equnu2}
\end{equation}
for $\nu=2$, where the pre-factors are fixed by the boundary conditions - no scattering into inner modes at $l=0$ and full equilibration at $l \gg l_r$.
Similarly, the TTC at filling fraction $\nu=4/3$ and $5/3$ can be written as \cite{Maiti2020}
\begin{equation}
^{4/3}G_{S2 \rightarrow D1} = \frac{1}{4}[3+ e^{-4l/l_r}] 
\label{equnu4b3}
\end{equation}   
and
\begin{equation}
^{5/3}G_{S2 \rightarrow D1} = \frac{1}{5}[3+ 2e^{-5l/2l_r}],
\label{equnu5b3}
\end{equation}
respectively. The above equations are utilized to estimate the value of equilibration rate $1/l_r$ from the measured TTC between the contacts $S2$ to $D1$ in Figs. \ref{fig:temp}(a), \ref{fig:temp}(b) and \ref{fig:temp}(c) (red curves).  It is assumed that the small amount of current that backscatters from the inner mode do not alter the above relations.

\begin{figure}[h!]
\includegraphics[width=7.5 cm]{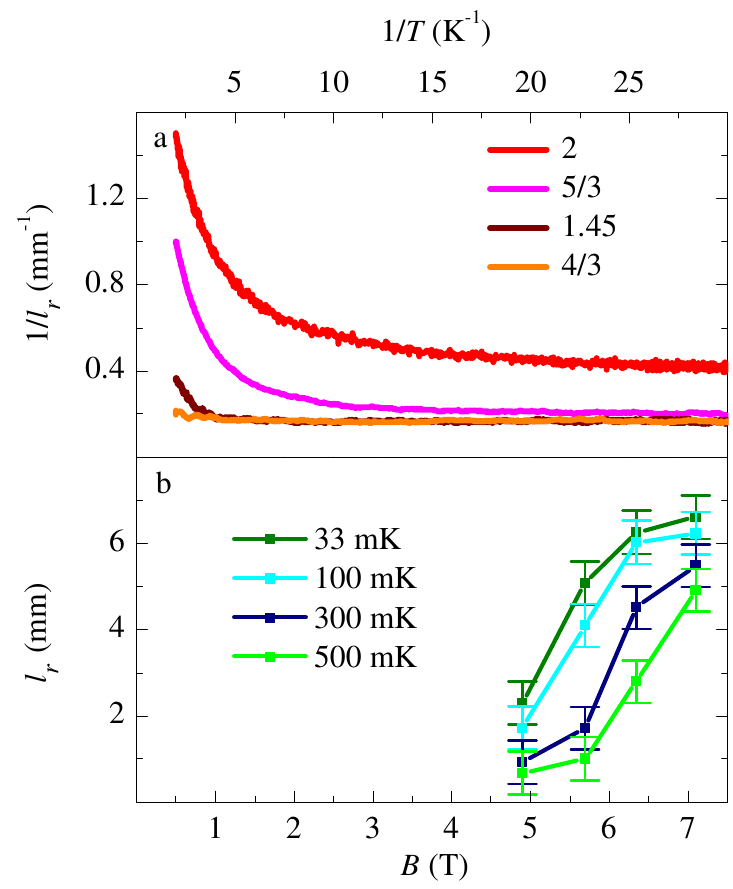}
\centering \caption[ ] 
{\label{fig:equilibration} (a)  Equilibration rate of the outer unity conductance mode versus $1/T$ at $\nu=4/3,1.45, 5/3$ and $2$. (b) Plot of equilibration length versus $B$ for different temperatures.}
\end{figure}

Equilibration rate $1/l_r$ of the outer integer mode is plotted as a function of $1/T$ in Fig. \ref{fig:equilibration}(a) for filling fraction $\nu = 4/3,\,5/3$ and $2$. At filling fraction $\nu = 1.45$ the QH state is compressible where the inner mode does not exist; however, an effective equilibration rate can be estimated for the outer mode using a similar exponential formulation and this shows an intermediate value as shown in Fig. \ref{fig:equilibration}(a). Equilibration rate for all the filling fractions increase monotonically with increasing temperature. The equilibration rates for fractional filling $\nu$ = 4/3 and 5/3 are similar at lowest temperature, but distinctly lower from that of integer filling $\nu$ = 2. To understand this observation at lowest temperature, the equilibration length of the outer unity conductance mode is estimated $6.6\pm0.5$ mm, $6.3\pm0.5$ mm, $5.1\pm0.5$ mm and $2.3\pm0.5$ mm  for filling fraction $\nu = 4/3, 1.45, 5/3$ and $2$, respectively and is plotted with magnetic field in Fig. \ref{fig:equilibration}(b). The equilibration length $l_r$ tends to saturate with increasing magnetic field (i.e. lowering filling fraction $\nu$) at lowest temperature and with increasing temperature the saturating trend of $l_r$ disappears [Fig. \ref{fig:equilibration}(b)].

\section{DISCUSION}
Edge mode structures of the QH states along smooth boundaries arise from formation of a sequence of dominant incompressible states as the electron density changes from the bulk value to zero at the boundary \cite{Beenakker1990,Chklovskii1992}. Here, two spatially separated edge modes are formed by incompressibility due to spin gap, and spin orthogonality condition prevents equilibration between these modes at lowest temperature. Therefore equillibration process requires a spin flip mechanism that is mediated by dynamic nuclear polarization \cite{Machinda2002,Yang2017,Kane1992,Wald1994,Dixon1997}.

Spin polarization of the QH system increases with lowering bulk filling fraction $\nu$ below 2 and the QH system become fully spin polarized at filling $\nu = 1$. This change of the spin polarization from spin un-polarized ($\nu = 2$) to spin polarized ($\nu = 1$) results in exchange enhancement of $g$-factor with lowering filling fraction \cite{Dolgopolov1997,Werner2020}. The enhanced spin gap is reflected in the observation of increase of threshold voltage $V_{th}$ for inter-mode transport with lowering filling fraction $\nu = 2$ to $4/3$ \cite{Deviatov2008}. The exchange enhanced spin gap induces spatial separation of the opposite spin modes. As a consequence of the increase in spatial separation, the equilibration length $l_r$ should increase with increasing magnetic field without showing saturation \cite{Maiti2020}. In contrast, the measured equilibration length $l_{r}$ of the outer unity conductance mode tends to saturate with increasing magnetic field $B$ as shown in Fig. \ref{fig:equilibration}(b) at lowest temperature.

In addition to the spin orthogonality, equilibration process may also be suppressed due to differing character of the electron like quasiparticles in the outer unity conductance mode and the anyon like quasiparticles in the correlated inner mode of $\nu = 5/3$ and 4/3 \cite{PhysRevB.67.045303}. This could explain the significantly larger equilibration rates in the integer QH state at $\nu = 2$ as compared to the fractional QH states at asymptotically low temperatures. A quantitative modeling of the observed saturating nature of equilibration length $l_r$ is left for further investigations.

\section{CONCLUSION}
In conclusion, we study equilibration between pairs of co-propagating edge modes of conductances $1$ on the outer side and $\nu-1$ on the inner side for bulk filling fraction $\nu = 2, 5/3$ and 4/3. We observe saturating nature of equilibration length $l_{r}$ of the outer unity conductance mode with increasing magnetic field at the lowest temperature. We argue that the significantly larger equilibration length for filling $\nu = 5/3$ and 4/3 compared to filling $\nu = 2$ is arising from suppression of equilibration due to differing character of the electron like quasiparticles in the outer unity conductance mode and the anyon like quasiparticles in the correlated inner mode.

\begin{center}
\textbf{ACKNOWLEDGEMENTS}
\end{center}                          
GJS acknowledges support from DST-SERB grant ECR/2018/001781. Part of the discussions in this and related previous works were facilitated by the International Centre for Theoretical Sciences program on Edge dynamics in topological phases (ICTS/edytop2019/06). SD acknowledges support from DST-SERB Grant No. MTR/2019/001043 and ARF grant from IISER Kolkata.

\begin{center}
\textbf{APPENDIX A}
\end{center}
In our experiments two terminal conductance (TTC) $^{\nu}G_{S \rightarrow D}$ is measured in a multi-terminal device as shown in Fig. \ref{fig:device diagram}(a). The measured TTC  $^{\nu}G_{S \rightarrow D}$ can be expressed in terms of transmission probabilities of the co-propagating edge modes connecting the source and detector. For this representation the outer mode and inner mode are labeled 1 and 2, respectively, where the outer mode has conductance unity and the inner mode has conductance 1, 2/3 and 1/3 for bulk filling fraction $\nu = 2, 5/3$ and 4/3, respectively. Following B\"{u}ttiker approach \cite{Buttiker1988,Beenakker1990,Maiti2020} the TTC for bulk filling fraction $\nu$ can be expressed as 
\begin{equation}
^{\nu}G_{\rm S \rightarrow D} = \sum_{i;j}{g_{i }} T_{ij}
\label{equap1}  
\end{equation} 
where $T_{ij}$ represents the probability of transmission from mode $i$ connected to the source $S$ into mode $j$ connected to the detector $D$. The summation $i(j)$ is taken over all the modes connected to the source (detector). $g_{i}$ is conductance of the $i$-th mode connected to the source $S$. Transmission probability $T_{ij}$ between the respective modes depend on the charge equilibration over the co-propagation length  $l = 125~\mu$m. In this presentation we exclude back scattered current reaching to detector $D6$.

Now we focus for the bulk filling fraction $\nu=4/3$, where the TTCs can be expressed as
\begin{equation}
^{4/3}G_{S2 \rightarrow D1} = T_{11},
\label{equap2}
\end{equation}
\begin{equation}
^{4/3}G_{S2 \rightarrow D2} = T_{12},
\label{equap3}
\end{equation}
\begin{equation}
^{4/3}G_{S1 \rightarrow D2} = \frac{1}{3}T_{22},
\label{equap4}
\end{equation} 
\begin{equation}
^{4/3}G_{S1 \rightarrow D1} = \frac{1}{3}T_{21}.
\label{equap5}
\end{equation} 
At lowest temperature, the two co-propagating modes do not equilibrate, hence the values of transmission probabilities become $T_{11}=T_{22}=1, T_{12}=T_{21}=0$. With increasing temperature, the two co-propagating mode starts equilibration into each other. If they fully equilibrate, the transmission probabilities become $T_{11}=3/4, T_{12}=1/4, T_{22}=1/4, T_{21}=3/4$. In our experiment temperature dependence of the TTCs are plotted in Fig. \ref{fig:temp} (a) and Fig. \ref{fig:temp} (d) for filling fraction $\nu=4/3$. Equilibration rate $1/l_{r}$ of the outer unity conductance mode for $\nu = 4/3$ is calculated using Eq. \ref{equnu4b3} from the measured value of $T_{11}$ [Eq. \ref{equap2}]. Similarly, TTCs at other filling fraction $\nu = 5/3$ and 2 can also be expressed.
\begin{center}
\textbf{APPENDIX B}
\end{center} 
In general, insulating states at the bulk gap melt at higher temperatures and start conducting, resulting in back scattering of the edge modes as depicted in Fig. \ref{fig:device diagram} (a) (orange dashed lines). In our device, back scattered current reaches the contact $D6$ when source $S1$ is excited and corresponding TTC $^{\nu}G_{S1 \rightarrow D6}$ increases with increasing temperature as shown in Fig. \ref{fig:temp}(d) for $\nu = 4/3$ and in Fig. \ref{fig:temp}(e) for $\nu = 5/3$ (orange curves). In a Hall bar device, the characteristic bulk gap is estimated from the Arrhenius plot of the finite longitudinal resistivity at elevated temperatures. A similar activation behavior is seen in our temperature dependent measurement [Fig. \ref{fig:temp}(d) and \ref{fig:temp}(e)], in which TTC from $S1$ to $D6$ can be expressed as \cite{Boebinger1987,Boebinger1985,Willett1988}
\begin{equation}
^{\nu}G_{S1 \rightarrow D6} \propto e^{-\frac{\Delta_{\nu}}{2T}},
\label{breakdown}
\end{equation}
where $\Delta_{\nu}$ is the bulk energy gap (also call pair creation energy) at filling fraction $\nu$. 
\begin{figure}[h!]
\includegraphics[width=7.5 cm]{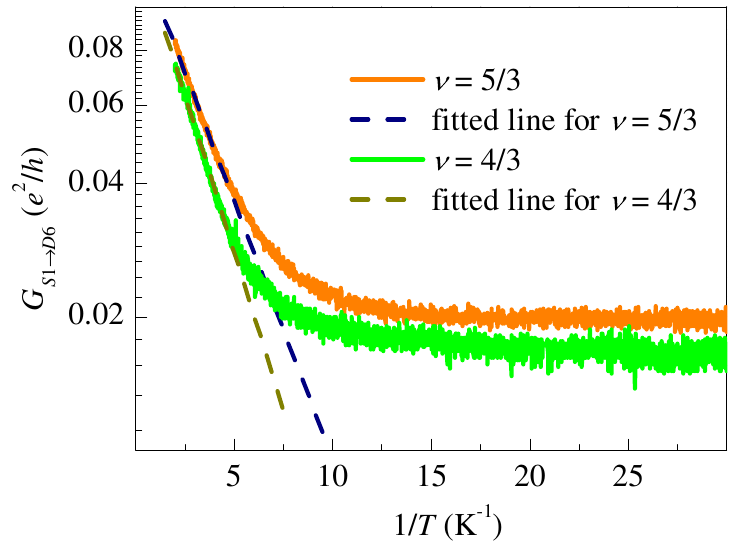}
\centering \caption[ ] 
{\label{fig:bulkgap} Arrhenius plot of the back-scattered conductance at filling $\nu$ = 4/3 and 5/3.}
\end{figure} 
The back-scattered conductance ($S1$ to $D6$) for filling fraction $\nu=4/3$ and $5/3$  are presented in [Fig. \ref{fig:bulkgap}]. The high temperature part of the data is fitted linearly to estimate the bulk gap and is found to be $\Delta_{4/3}=0.652 \pm 0.06$K and $\Delta_{5/3}=0.534 \pm 0.06$K for filling fraction $\nu=4/3$ and $5/3$, respectively. The measured bulk gaps are consistent with the previous measurements \cite{Boebinger1987,Boebinger1985}.

\bibliography{reference}
\end{document}